\documentclass[runningheads]{llncs}

\usepackage{graphicx}
\usepackage{hyperref}
\usepackage{textcomp}
\usepackage{bbding}
\usepackage[font=scriptsize,labelfont=bf]{caption}
\usepackage{xcolor}
\usepackage{pgf}
\usepackage{tikz}
\usetikzlibrary{arrows, automata, shapes, petri, positioning, calc}
\usepackage[export]{adjustbox}

\hypersetup{
	colorlinks=true,
	linkcolor={red!50!black},
	citecolor={blue!60!black},
	urlcolor={blue!80!black},
	pdfauthor={Benevento Elisabetta, Pegoraro Marco, Mattia Antoniazzi, Beyel Harry H., Peeva Viki, Balfanz Paul, van der Aalst Wil M.P., Martin Lukas, Marx Gernot},
	pdftitle={{Process Modeling and Conformance Checking in Healthcare: A COVID-19 Case Study}},
	pdfsubject={Process Mining in Healthcare},
	pdfkeywords={Process Mining, Healthcare, COVID-19, STAKOB Guidelines, Business Process Management, Conformance Checking},
	pdfproducer={LaTeX},
	pdfcreator={pdfLaTeX},
	bookmarksopen=true
}

\begin{document}

\title{Process Modeling and Conformance Checking in Healthcare: A COVID-19 Case Study\thanks{
		We acknowledge the ICU4COVID project (funded by European Union's Horizon 2020 under grant agreement n. 101016000) and the COVAS project.
}}
\subtitle{Case Study}

\titlerunning{Process Modeling and Conformance Checking in Healthcare}

\author{Elisabetta Benevento\,\Envelope\,\inst{1, 2}\orcidID{0000-0002-3999-8977} \and Marco Pegoraro\,\Envelope\,\inst{1}\orcidID{0000-0002-8997-7517} \and Mattia Antoniazzi\inst{2}\orcidID{0000-0001-9467-1988} \and Harry H. Beyel\inst{1}\orcidID{0000-0002-6541-3848} \and Viki Peeva\inst{1}\orcidID{0000-0001-7144-5136} \and Paul Balfanz\inst{3}\orcidID{0000-0001-9539-4804} \and Wil M.P. van der Aalst\inst{1}\orcidID{0000-0002-0955-6940} \and Lukas Martin\inst{4}\orcidID{0000-0001-8650-5090} \and Gernot Marx\inst{4}}

\authorrunning{Benevento et al.}

\institute{Chair of Process and Data Science (PADS),\\Department of Computer Science, RWTH Aachen University, Aachen, Germany
	\email{\{benevento, pegoraro, beyel, peeva, vwdaalst\}@pads.rwth-aachen.de}\\
	\and
    Department of Energy, Systems, Territory and Construction Engineering, University of Pisa, Pisa, Italy
    \and
    Department of Cardiology, Angiology and Intensive Care Medicine,\\RWTH University Hospital, Aachen, Germany
    \email{\\pbalfanz@ukaachen.de}\\
    \and
	Department of Intensive Care and Intermediate Care,\\RWTH Aachen University Hospital, Aachen, Germany
	\email{\\\{lmartin, gmarx\}@ukaachen.de}
}

\maketitle

\begin{abstract}

The discipline of process mining has a solid track record of successful applications to the healthcare domain. Within such research space, we conducted a case study related to the Intensive Care Unit (ICU) ward of the Uniklinik Aachen hospital in Germany. The aim of this work is twofold: developing a normative model representing the clinical guidelines for the treatment of COVID-19 patients, and analyzing the adherence of the observed behavior (recorded in the information system of the hospital) to such guidelines. We show that, through conformance checking techniques, it is possible to analyze the care process for COVID-19 patients, highlighting the main deviations from the clinical guidelines. The results provide physicians with useful indications for improving the process and ensuring service quality and patient satisfaction. We share the resulting model as an open-source BPMN file.


\keywords{Process Mining \and Healthcare \and COVID-19 \and STAKOB Guidelines \and Business Process Management \and Conformance Checking.}
\end{abstract}

\setcounter{footnote}{0}

\section{Introduction}
At the turn of the decade, the logistics of operations in hospitals and healthcare centers have been severely disrupted worldwide by the COVID-19 pandemic. Its impact has been profound and damaging in all aspects of life, but in no context it has been more damaging than in healthcare: the safety and well-being of physicians and medical personnel, the supply chain of drugs and equipment, and the capacity of hospitals were all challenged by the pandemic. 

One of the most critical points for healthcare systems involved in the treatment process is the management of COVID-19 patients needing acute and respiratory care. Therefore, healthcare organizations are increasingly pushed to improve the efficiency of care processes and the resource management for such category of patients. One way to attain such improvement is to leverage historical data from information systems of hospitals. These data can be then cleaned and analyzed, to individuate non-compliant behavior and inefficiencies in the care process.

The aim of our work is to analyze the care process for the COVID-19 patients treated at the Intensive Care Unit (ICU) ward of the Uniklinik Aachen hospital in Germany, in order to identify divergences or anomalies within the process. To do so, our work intends to develop an executable process model representing the clinical guidelines for the treatment of COVID-19 patients and evaluate the adherence of the observed behavior (recorded by the information system of the hospital) to such guidelines.

The STAKOB guidelines\footnote{\url{https://www.rki.de/DE/Content/Kommissionen/Stakob/Stakob_node.html}} (``Ständigen Arbeitskreis der Kompetenz- und Behandlungszentren für Krankheiten durch hochpathogene Erreger'', ``Permanent working group of competence and treatment centers for diseases caused by highly pathogenic agents'') are widely accepted and recognized protocols for the treatment of COVID-19, compiled and verified by a large consensus of medical scientists, physicians, and research institutions. They provide a comprehensive overview of recommendations on the management of hospitalized COVID-19 patients. The process model was obtained starting from such guidelines, and was validated by the physicians working in the intensive and intermediate care unit of the Uniklinik. We openly share the resulting BPMN model, as well as the related documentation. The conformance with the guidelines was assessed by using process mining techniques. The results provide hospital managers with information about the main deviations and/or anomalies in the process and their possible causes. In addition, they suggest improvements to make the process more compliant, cost-effective, and performant.

The remainder of the paper is structured as follows. Section~\ref{sec:related} explores related work and sets the context of our research. Section~\ref{sec:method} lays out the methodology we employed in our case study. Section~\ref{sec:results} illustrates the results of our case study. Finally, Section~\ref{sec:conclusion} concludes the paper.

\section{Related Work}\label{sec:related}

The global effort to fight the pandemic has stimulated the adoption of new technologies in healthcare practice~\cite{golinelli2020adoption}. An area where this effect has been radical is the digitization of healthcare processes, both medical and administrative. Data recording and availability have improved during the years of the pandemic. Stakeholders realized that data are a valuable information source to support the management and improvement of healthcare processes~\cite{munoz2022process}. In addition, the reliance of medical personnel on digital support systems is now much more significant.
Fields of science that have recently shown to be particularly promising when applied to healthcare operations are the process sciences, and specifically Business Process Management (BPM) and process mining~\cite{munoz2022process}. This is mainly due to the characteristics of healthcare process, which are complex and flexible and involve a multidisciplinary team~\cite{munoz2022process,rebuge2012business}. Particularly, process mining has emerged as a suitable approach to analyze, discover, improve, and manage real-life and complex processes, by extracting knowledge from event logs~\cite{van2016process}. Currently, process scientists have gathered event data on the process of treatment for COVID-19 and leveraged process mining techniques to obtain insights on various aspects of the healthcare process~\cite{pegoraro2022analyzing,augusto2022process,dos2021process} or on how other business processes have been impacted by the disruption caused by COVID-19~\cite{DBLP:conf/bpm/ZabkaBA21}.

Among process mining techniques, conformance checking aims to measure the adherence of a (discovered or known) process with a given set of data, or vice-versa~\cite{gatta2019clinical}. Conformance checking helps medics to understand major deviations from clinical guidelines, as well as to identify areas for improvement in practices and protocols~\cite{munoz2022process}. Some studies have applied these techniques in different healthcare contexts, such as oncology~\cite{rojas2016process}. However, no studies have addressed the compliance analysis on the care process of COVID-19 patients in a real-life scenario. 
To do so, it is essential to have a normative model, reflecting clinical guidelines and protocols, that can be interpreted by machines. Currently, executable process models representing the guidelines for the treatment of COVID-19 patients are still absent and needed, given the uncertainty and variability of the disease.

\section{Methodology}\label{sec:method}

\begin{figure}[t]
\centering
\includegraphics[width=\textwidth]{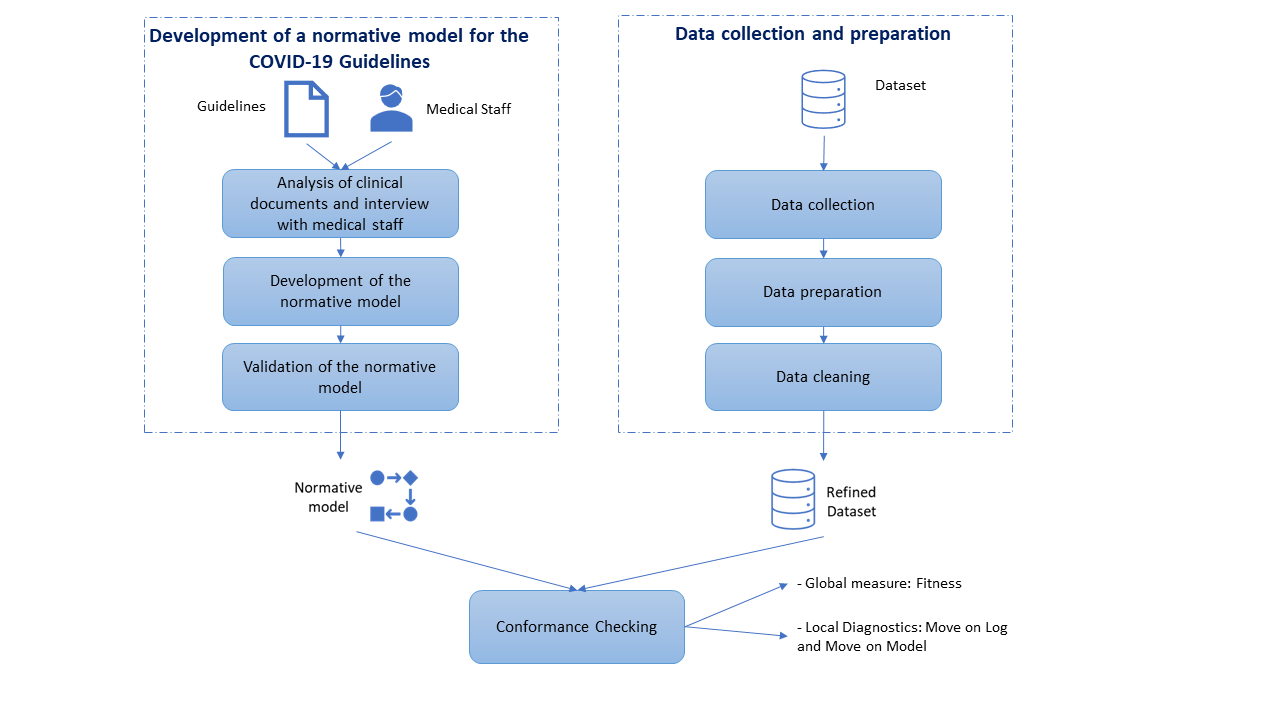}
\caption{Case study methodology. Our work measures the deviation between the expected and real behavior of the COVID-19 treatment process, respectively represented by the STAKOB guidelines, and by the COVAS dataset.}
\label{fig:metodo}
\end{figure}

The methodology conducted in this study consists of the following three main steps, also shown in Figure~\ref{fig:metodo}:
\begin{itemize}
\item Development of a normative model based on the STAKOB guidelines. A normative model is a process model that reflects and implements rules, guidelines, and policies of the process, mandated by process owners or other supervisory bodies. This phase involves (i) the analysis of the STAKOB documentation and interview with ICU physicians, (ii) the development of the model from the guidelines, and (iii) the validation of the model with ICU physicians. 
\item Data collection and preparation, which involves the extraction and preprocessing of event data, gathered from the information system of the hospital. The event log is refined by removing duplicate and irrelevant data, handling missing data, and detecting outliers to ensure data reliability.
\item Conformance checking, which involves the use of conformance checking techniques to compare the normative model with the event logs for the three COVID-19 waves and determine whether the behavior observed in practice conforms to the documented process.
\end{itemize}

\subsection{Development of a normative model based on the STAKOB guidelines}

The STAKOB guidelines provide information on the disease and its related symptoms, and describe the diagnostic and treatment activities to be performed on COVID-19 patients and the therapies to be administered. The treatment of COVID-19 patients requires a multi-disciplinary approach: in addition to intensive care physicians and nurses, specialists in infectious diseases and infection control must also be part of the team~\cite{malin2022key}. The guidelines guide the operations of the medical team involved in the inpatient care of COVID-19 patients, but are also intended to provide information for individuals and/or organizations directly involved in this topic. 

To make the guidelines interpretable by machines---and thus suitable for conformance checking---we developed a normative process model of the STAKOB guidelines in the BPMN language using the Signavio tool\footnote{\url{https://www.signavio.com/}}. The choice of the BPMN standard is due to its ability to be executable but, at the same time, easy to understand by physicians and practitioners.
The BPMN model of the STAKOB guidelines was validated by using a qualitative approach. Specifically, the model was presented and discussed with three physicians working in the intensive and intermediate care unit of the Uniklinik during three meetings. During the meetings, several refinements were applied to the model, until it was approved by all.

\subsection{Data Collection and Preparation}
We collected and pre-processed data of COVID-19 patients monitored in the context of the COVID-19 Aachen Study (COVAS). The log contains event information regarding COVID-19 patients treated by the Uniklinik between January 2020 and June 2021. Events (patient admittance, symptoms, treatments, drug administration) are labeled with the date, creating timestamps with a coarseness at the day level. While here we exclusively focus on process mining, the COVAS dataset has also been analyzed in the context of explainable AI~\cite{DBLP:conf/ideal/VeliogluGAH22}.

Data were gathered from the information system of the hospital. The initial database consisted of 269 cases, 33 activity labels, 210 variants, and 3542 events. Before the analysis, we refined the raw event log, to guarantee its quality.
Data cleaning and preparation were executed with Python and included: (i) outliers and incomplete cases removal based on the number of hospitalization days, (ii) less significant activities abstraction, and (iii) filtering of infrequent variants. As an example, we removed the cases with a duration of more than 70 days: this value was validated with the doctors, according to whom durations longer than 70 days may be due to registration delays. In the end, the refined event log consisted of 187 patient cases, 32 activities, 135 variants, and 2397 events.

\begin{figure}[t]
\centering
\includegraphics[width=\textwidth]{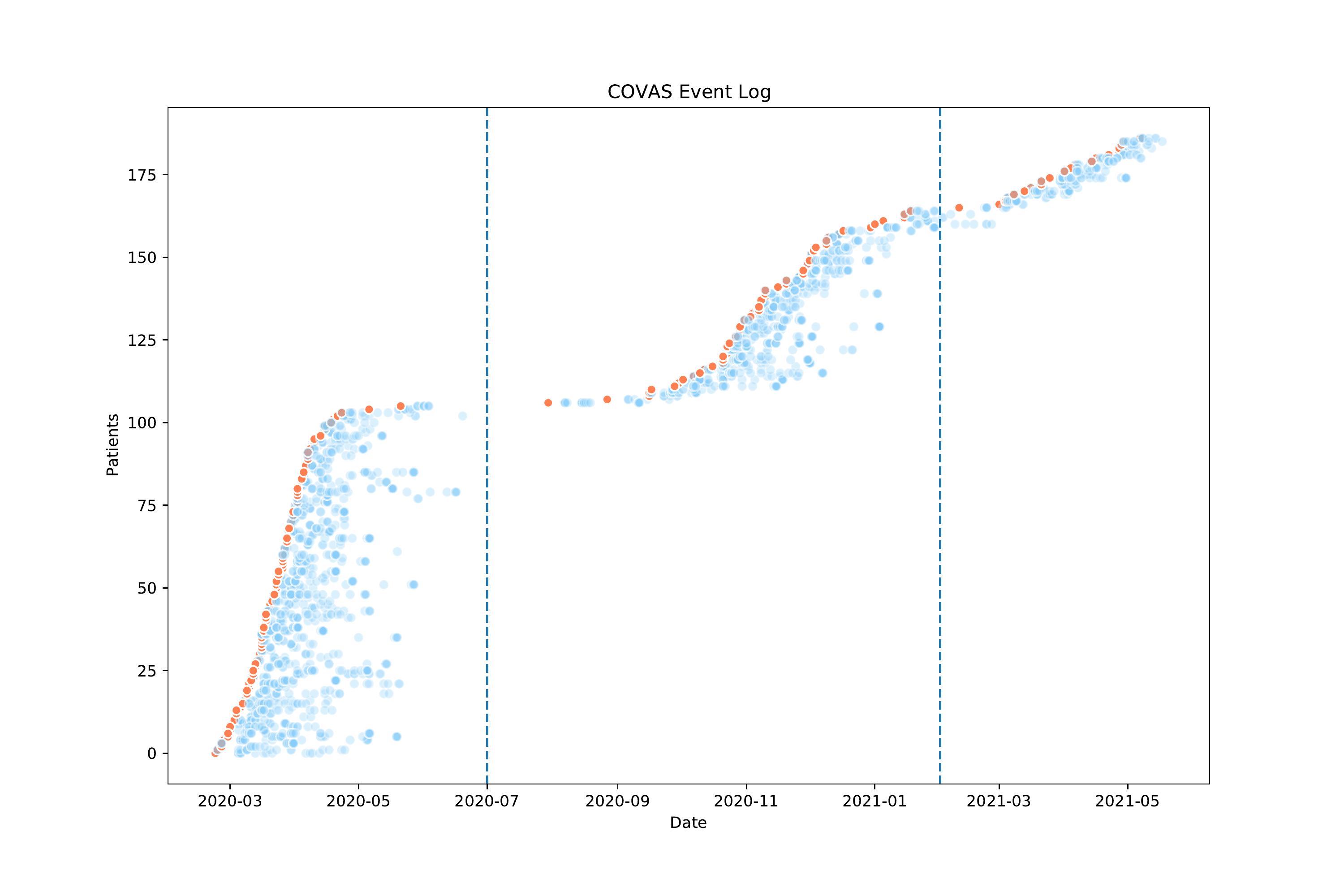}
\caption{Dotted chart of the COVAS event log. The cases are sorted by the first recorded event, which is highlighted in orange. Every blue dot corresponds to a recorded event. The vertical dashed lines separate the first, second, and third COVID-19 waves, based on the knowledge of physicians.}
\label{fig:3onde}
\end{figure}

To evaluate the adherence of the COVAS dataset to the normative model during the three COVID-19 waves, we split the dataset into three sub-event logs. As illustrated in the next sections, this is done with the goal of examining how treatment operations for COVID-19 change between infection waves with respect to the adherence to the STAKOB guidelines. As shown by the dotted chart of the event log in Figure~\ref{fig:3onde}, the three waves can be clearly identified. Such a choice of wave separation was also supported by the literature~\cite{dongelmans2022characteristics}.

The event log of the first wave contains 106 cases and 1410 events. The average duration of the process is 25.38 days. The log of the second wave contains 59 cases and 892 events, with an average duration of 22.42 days. The log of the third wave contains 22 cases and 282 events, with an average duration of 16.38 days.

\subsection{Conformance Checking}

For each sub-event log, we applied conformance checking techniques to identify deviations within the process. Specifically, we utilized the plug-in ``Replay a Log on Petri Net for Conformance Analysis'' as implemented on ProM, with standard setting parameters. The choice is due to the fact that alignment-based techniques can exactly pinpoint where deviations are observed~\cite{van2016process,adriansyah2010towards}.  

The alignment-based technique allowed to estimate a global conformance measure, which quantifies the overall conformance of the model and event log, and local diagnostics, which identify points where the model and event log do not agree. In the first case, we calculated fitness, which measures ``the proportion of behavior in the event log possible according to the model''~\cite{van2016process}. In the second case, we estimated for each activity within the model the following~\cite{dixit2017enabling}:
\begin{itemize}
    \item the number of ``moves on log'': Occurrences of an activity in the trace cannot be mapped to any enabled activity in the process model.
    \item the number of ``moves on model'': Occurrences of an enabled activity in the process model cannot be mapped to any event in the trace sequence.
    \item the number of ``synchronous moves'': Occurrences of an activity belonging to a trace can be mapped to occurrences of an enabled activity in the process model. 
\end{itemize}

\section{Results}\label{sec:results}
In this section, we presented the results from the development of the normative model and the conformance checking analysis.

\subsection{Normative Model}
\begin{figure}[t]
\centering
\includegraphics[width=\textwidth]{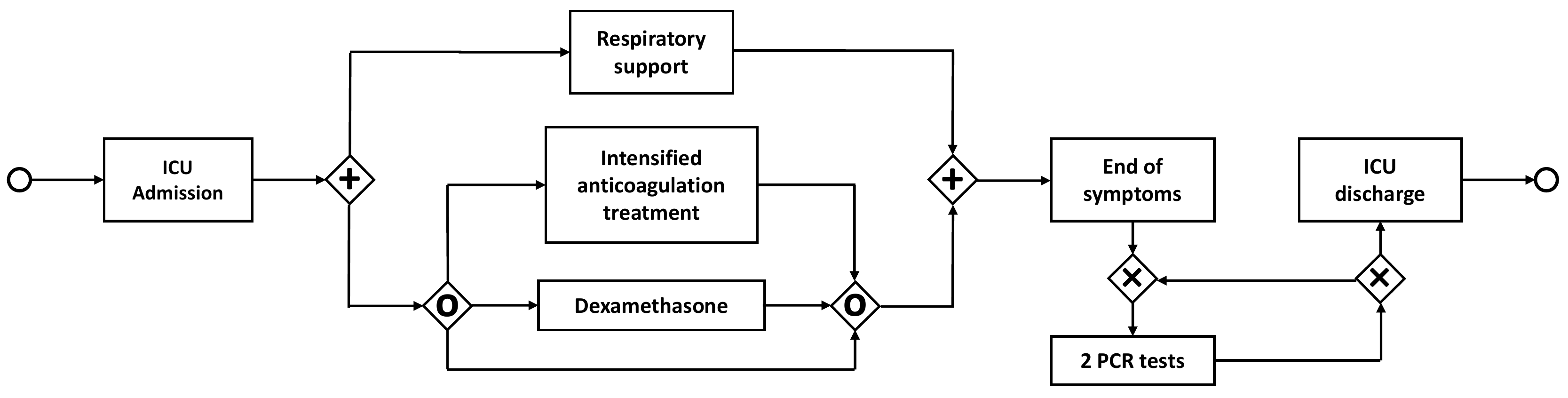}
\caption{A section of the STAKOB COVID-19 model, depicting some activities related to the ICU operations for COVID-19 patients.}
\label{fig:stakob1}
\end{figure}

The developed normative model consists of 3 sub-processes, 23 activities and approximately 36 gateways (XOR, AND and OR). Figure~\ref{fig:stakob1} shows a section of the model. 

\begin{figure}[t]
\centering
\includegraphics[width=\textwidth]{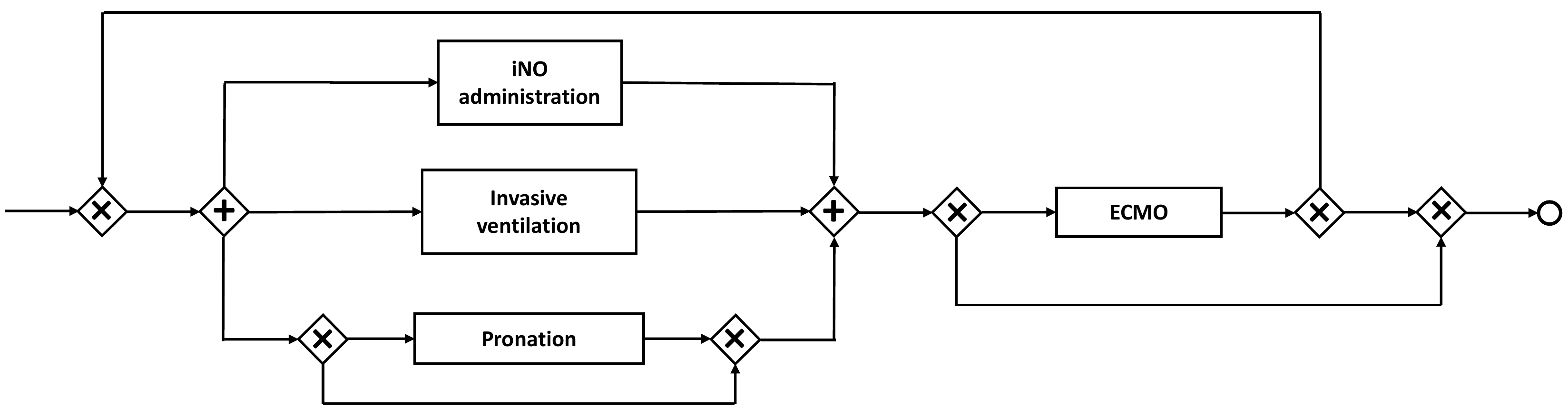}
\caption{A section of the STAKOB COVID-19 model, depicting some activities related to the respiration support operations for COVID-19 patients.}
\label{fig:stakob2}
\end{figure}

The model clearly underlines the fact that the treatment of hospitalized patients with COVID-19 is complex and is characterized by several pursuable pathways (see the presence of XOR and OR gateways). It also requires the collaboration of different departments and specialists. More in detail, the care treatment includes an antibiotic/drug therapy phase and, if necessary, an oxygenation phase. At this point, if the patient’s health condition deteriorates, the transfer to the ICU is planned (partially shown in Figure~\ref{fig:stakob1}). In the ICU, the patient may undergo mechanical ventilation, ECMO (ExtraCorporeal Membrane Oxygenation) or pronation in addition to the medical therapy. A section of the sub-process showing the respiratory support for the patient can be seen in Figure~\ref{fig:stakob2}. Recovery and subsequent discharge are confirmed by two negative COVID-19 tests.

The full model is openly available on GitHub\footnote{\url{https://github.com/marcopegoraro/pm-healthcare/tree/main/stakob}}. It is rendered in the XML export format of the BPMN standard\footnote{\url{https://www.bpmn.org/}}. The folder also contains a PDF depicting the entire model, a license declaration, and an addendum describing the model schematic in more detail.

\subsection{Conformance Checking Results}

\subsubsection{COVID-19 First Wave Results}

For the first wave, the fitness between the model and the data is 0.69; some trace variants are not reproduced by the model. This may be due to the variability of the process (health conditions vary from patient to patient). In addition, the coarseness of the timestamps in the dataset has an impact: events are recorded at the date level, so the order in which they are recorded may vary in some instances.
Table~\ref{tab:wave1} shows the results of the conformance checking for the first wave. Specifically, for each activity, it shows the misalignments between the normative model and the event log.

\begin{table}[t]
\centering
\caption{Results of conformance checking alignments with the STAKOB model for the patient sub-log corresponding to the first COVID-19 wave. For each activity in the log, we show the count of moves on log, moves on model, and synchronous moves.}
\label{tab:wave1}
\scriptsize
\begin{tabular}{|l|c|c|c||l|c|c|c|}
\hline
\textbf{Activity} & \textbf{\vbox{\hbox{Move on}\hbox{log}}} & \textbf{\vbox{\hbox{Syncro}\hbox{move}}} & \textbf{\vbox{\hbox{Move on}\hbox{model}}} & \textbf{Activity} & \textbf{\vbox{\hbox{Move on}\hbox{log}}} & \textbf{\vbox{\hbox{Syncro}\hbox{move}}} & \textbf{\vbox{\hbox{Move on}\hbox{model}}} \\ \hline
Symptobegin       & 0                    & 106                  & 0                      & Ventilation Start & 33                   & 9                    & 2                      \\ \hline
Hospitalization   & 1                    & 105                  & 1                      & Ventilation End   & 35                   & 8                    & 6                      \\ \hline
UKA Admission     & 12                   & 96                   & 10                     & NMB Start         & 4                    & 11                   & 0                      \\ \hline
Abx Start         & 2                    & 58                   & 0                      & NMB End           & 4                    & 11                   & 0                      \\ \hline
Abx End           & 2                    & 58                   & 0                      & CVVH Start        & 16                   & 11                   & 0                      \\ \hline
Start Oxygen      & 22                   & 85                   & 0                      & CVVH End          & 16                   & 11                   & 0                      \\ \hline
Remdesivir Start  & 0                    & 3                    & 0                      & Prone Start       & 25                   & 10                   & 0                      \\ \hline
Remdesivir End    & 0                    & 3                    & 0                      & Prone End         & 25                   & 10                   & 0                      \\ \hline
Admission ICU     & 35                   & 20                   & 0                      & ECMO Start        & 10                   & 0                    & 0                      \\ \hline
HiFlo Start       & 0                    & 1                    & 19                     & ECMO End          & 10                   & 0                    & 0                      \\ \hline
Hiflo End         & 0                    & 1                    & 19                     & End Of Fever      & 22                   & 53                   & 53                     \\ \hline
NIV Start         & 6                    & 5                    & 9                      & Discharge ICU     & 48                   & 6                    & 14                     \\ \hline
NIV End           & 10                   & 5                    & 9                      & Last Oxygen Day   & 39                   & 53                   & 53                     \\ \hline
iNO Start         & 13                   & 10                   & 1                      & Discharge dead    & 0                    & 33                   & 0                      \\ \hline
iNO End           & 13                   & 10                   & 1                      & Discharge alive   & 0                    & 73                   & 0                      \\ \hline
\end{tabular}
\normalsize
\end{table}

Several misalignments can be observed. In particular:
\begin{itemize}
    \item The \emph{HiFlo Start} and \emph{HiFlo End} activities (corresponding to high flow oxygenation) present 19 moves on model and one synchronous move. This means that, although it is required by the guidelines, this activity is only performed in one case. This indicates that, given the patient's condition, the physicians may have seen fit to skip this treatment.
    \item There are several tasks that have both moves on model and moves on log. This means that these tasks often deviate from the normative model (in some cases they are present in the model but not in reality, in others vice-versa). This may be due to the variability of patients' conditions and the lack of familiarity with COVID-19 and its standardized treatment, since this data was recorded in the early days of the pandemic. For example, the guidelines suggest that the \emph{Discharge ICU} should occur after ventilation and pronation, while in reality, in some cases, it occurs before. Thus, many activities occur while the patient is hospitalized, but not still formally admitted to the ICU. 
    \item Some activities present only moves on log and synchronous moves, i.e., they are present in reality but at times not in the normative model. This means that they are performed at different times than the guidelines suggest. For example, \emph{Admission ICU} may be anticipated because of a particularly critical course not foreseen by the physicians or be delayed because no space in ICU is available at that time; or \emph{Prone End} (the interruption of the treatment of pronation) may be brought forward because of the negative effects on the patient, e.g., the appearance of pressure sores. Alternatively, pronation may be delayed because the patient has not achieved optimal arterial blood oxygenation.
\end{itemize}

\subsubsection{COVID-19 Second Wave Results}

For the log of the second wave, the fitness with the STAKOB model is 0.66. 
Table~\ref{tab:wave2} shows the results of conformance checking for the second wave. 

\begin{table}[t]
\centering
\caption{Results of conformance checking alignments with the STAKOB model for the patient sub-log corresponding to the second COVID-19 wave. For each activity in the log, we show the count of moves on log, moves on model, and synchronous moves.}
\label{tab:wave2}
\scriptsize
\begin{adjustbox}{width=\columnwidth, center}
		\begin{tabular}{|l|c|c|c||l|c|c|c|}
			\hline
			\textbf{Activity} & \textbf{\vbox{\hbox{Move on}\hbox{log}}} & \textbf{\vbox{\hbox{Syncro}\hbox{move}}} & \textbf{\vbox{\hbox{Move on}\hbox{model}}} & \textbf{Activity} & \textbf{\vbox{\hbox{Move on}\hbox{log}}} & \textbf{\vbox{\hbox{Syncro}\hbox{move}}} & \textbf{\vbox{\hbox{Move on}\hbox{model}}} \\ \hline
			Symptobegin         & 0                    & 59                   & 0                      & Dexamethasone End & 24                   & 14                   & 1                      \\ \hline
			Hospitalization     & 0                    & 59                   & 0                      & Ventilation Start & 11                   & 8                    & 1                      \\ \hline
			UKA Admission       & 8                    & 50                   & 9                      & Ventilation End   & 11                   & 8                    & 1                      \\ \hline
			Abx Start           & 0                    & 29                   & 0                      & NMB Start         & 2                    & 9                    & 0                      \\ \hline
			Abx End             & 0                    & 29                   & 0                      & NMB End           & 2                    & 9                    & 0                      \\ \hline
			Start Oxygen        & 5                    & 54                   & 0                      & CVVH Start        & 7                    & 8                    & 1                      \\ \hline
			Remdesivir Start    & 8                    & 12                   & 0                      & CVVH End          & 7                    & 8                    & 1                      \\ \hline
			Remdesivir End      & 8                    & 12                   & 0                      & Prone Start       & 8                    & 8                    & 0                      \\ \hline
			Admission ICU       & 8                    & 15                   & 1                      & Prone End         & 8                    & 8                    & 0                      \\ \hline
			HiFlo Start         & 0                    & 2                    & 14                     & ECMO Start        & 7                    & 0                    & 0                      \\ \hline
			Hiflo End           & 0                    & 2                    & 14                     & ECMO End          & 7                    & 0                    & 0                      \\ \hline
			NIV Start           & 6                    & 8                    & 5                      & End Of Fever      & 27                   & 13                   & 43                     \\ \hline
			NIV End             & 8                    & 5                    & 8                      & Discharge ICU     & 20                   & 2                    & 14                     \\ \hline
			iNO Start           & 2                    & 9                    & 0                      & Last Oxygen Day   & 19                   & 36                   & 23                     \\ \hline
			iNO End             & 2                    & 9                    & 0                      & Discharge dead    & 0                    & 17                   & 0                      \\ \hline
			Dexamethasone Start & 23                   & 15                   & 0                      & Discharge alive   & 0                    & 42                   & 0                      \\ \hline
		\end{tabular}
	\end{adjustbox}
\normalsize
\end{table}

In the second wave, \emph{Hospitalization} is only performed after the onset of symptoms, as suggested by the guidelines.
However, deviations are also encountered. As in the first wave, the most affected activities are \emph{End Of Fever}, \emph{Admission ICU} and \emph{Discharge ICU}, and \emph{Last Oxygen Day}, which have both moves on log and moves on model. This may be related to the mutability of the disease becoming difficult to manage with common protocols and the variability of the patients' conditions. 
Compared to the first wave, the use of drugs has changed. In particular, a new drug is being administered, i.e., Dexamethasone, and the use of Remdesivir is increased. The administration of both drugs has moves on log mismatches, indicating that the physicians needed to administer such treatments more frequently than recommended. The former is also used in patients who do not require intensive care, contrary to what the guidelines suggest. The second, which is preferred for non-critical hospitalized patients, is also used in intensive care. 
In addition, high flow oxygenation is rarely performed here, despite being included in the guidelines. 

\subsubsection{COVID-19 Third Wave Results}

The fitness between the log and the model is 0.69 for the third COVID-19 wave.
Table~\ref{tab:wave3} shows the results of conformance checking for the third wave.

\begin{table}[]
\centering
\caption{Results of conformance checking alignments with the STAKOB model for the patient sub-log corresponding to the third COVID-19 wave. For each activity in the log, we show the count of moves on log, moves on model, and synchronous moves.}
\label{tab:wave3}
\scriptsize
\begin{adjustbox}{width=\columnwidth, center}
		\begin{tabular}{|l|c|c|c||l|c|c|c|}
			\hline
			\textbf{Activity} & \textbf{\vbox{\hbox{Move on}\hbox{log}}} & \textbf{\vbox{\hbox{Syncro}\hbox{move}}} & \textbf{\vbox{\hbox{Move on}\hbox{model}}} & \textbf{Activity} & \textbf{\vbox{\hbox{Move on}\hbox{log}}} & \textbf{\vbox{\hbox{Syncro}\hbox{move}}} & \textbf{\vbox{\hbox{Move on}\hbox{model}}} \\ \hline
			Symptobegin         & 0                    & 22                   & 0                      & Dexamethasone End & 8                    & 4                    & 0                      \\ \hline
			Hospitalization     & 2                    & 19                   & 3                      & Ventilation Start & 1                    & 9                    & 1                      \\ \hline
			UKA Admission       & 0                    & 22                   & 0                      & Ventilation End   & 1                    & 9                    & 1                      \\ \hline
			Abx Start           & 0                    & 8                    & 0                      & NMB Start         & 0                    & 1                    & 0                      \\ \hline
			Abx End             & 0                    & 8                    & 0                      & NMB End           & 0                    & 1                    & 0                      \\ \hline
			Start Oxygen        & 0                    & 38                   & 0                      & CVVH Start        & 2                    & 1                    & 0                      \\ \hline
			Remdesivir Start    & 0                    & 1                    & 0                      & CVVH End          & 2                    & 1                    & 0                      \\ \hline
			Remdesivir End      & 0                    & 1                    & 0                      & Prone Start       & 1                    & 1                    & 0                      \\ \hline
			Admission ICU       & 1                    & 2                    & 1                      & Prone End         & 1                    & 1                    & 0                      \\ \hline
			HiFlo Start         & 0                    & 2                    & 1                      & ECMO Start        & 0                    & 0                    & 0                      \\ \hline
			Hiflo End           & 0                    & 2                    & 1                      & ECMO End          & 0                    & 0                    & 0                      \\ \hline
			NIV Start           & 4                    & 1                    & 0                      & End Of Fever      & 11                   & 6                    & 16                     \\ \hline
			NIV End             & 5                    & 0                    & 1                      & Discharge ICU     & 3                    & 2                    & 1                      \\ \hline
			iNO Start           & 1                    & 1                    & 0                      & Last Oxygen Day   & 3                    & 17                   & 5                      \\ \hline
			iNO End             & 1                    & 1                    & 0                      & Discharge dead    & 0                    & 3                    & 0                      \\ \hline
			Dexamethasone Start & 9                    & 3                    & 1                      & Discharge alive   & 0                    & 19                   & 0                      \\ \hline
		\end{tabular}
	\end{adjustbox}
\normalsize
\end{table}

The physicians' experience and familiarity with the disease appear to have increased. However, many of the misaligned activities have similar behavior to those performed during past waves. 
Note that the ECMO treatment has zero values in all columns. This is because it is not performed in the third wave (unlike the first two). Since ECMO is the most invasive oxygenation treatment, this may be due to the fact that the severity of the patients' condition has decreased. 

To summarize, alignments-based techniques make it possible to detect and analyze process deviations, providing useful insights for physicians. Furthermore, in the three waves, most activities remained misaligned, while some moved closer to the guidelines' suggestion. This shows that the process is highly variable and specific care pathways are required for each patient, which do not always coincide with those stated in the guidelines. 

\section{Conclusion}\label{sec:conclusion}

Our work aimed to analyze the care process for COVID-19 patients, bringing to light deviations from the clinical guidelines. Specifically, the work proposed a normative model bases on the STAKOB guidelines, which can be interpreted by software tools (e.g., process mining software). The BPMN model is openly accessible to any analyst, and can also be loaded into any commercial software supporting the BPMN standard, like Celonis and Signavio. This addresses the need for computer-interpretable and usable guidelines in healthcare, particularly for the treatment of COVID-19 patients~\cite{oliart2022we}. In addition, the work provided physicians with guidance on the management of COVID-19 patients, highlighting deviations and critical points in the three infection waves.

The contributions of our work are: 
\begin{itemize}
    \item One of the first attempts to apply a process mining-based methodology for the analysis of process deviations in a real, complex, and uncertain healthcare context, like the recent and ongoing COVID-19 pandemic.
    \item The development of a normative model that can advise physicians in the treatment of COVID-19 patients by providing specific guidelines and procedures to follow. This is helpful in dealing with the uncertainty and complexity of healthcare operations brought about by the pandemic. In addition, the model can be used as input for the development of a decision support system, which alerts in real-time in case of violations of the guidelines.
    \item The extraction of valuable insights for physicians regarding the main deviations and the related causes in the COVID-19 patient care process. This knowledge is crucial for improving the process and ensuring service quality and patient satisfaction, e.g., better management of drug administration (when to administer and how often), more targeted execution of certain treatments---e.g., pronation---(who to treat and when to do it), and execution of treatments suggested by guidelines but never performed in reality that can enhance the care pathway and reduce hospitalization time (such as high flow oxygenation).
\end{itemize}

The work presents some open questions and directions for future research. The limited size, especially for the third wave, and the coarseness of the timestamps in the dataset may impact the results. To address this issue, a possible option is to weigh the results of analyses using the probability of specific orderings of events in traces~\cite{DBLP:conf/icpm/0001BUA21}. Furthermore, the physician's consensus on both the validity of the STAKOB model and the interpretation of the conformance checking results can definitely be enlarged, by soliciting the expert opinion of a larger group of medics.
As future developments, we plan to: (i) extend the research and collect new data from other German hospitals, in order to generalize the results and identify best practices in the treatment of COVID-19 patients; (ii) improve the validation of results; (iii) actively involve physicians in the analysis of deviations, using qualitative approaches such as interviews and field observations; (iv) conduct a more extensive comparative analysis based on process mining, including a structural model comparison, concept drift, and performance analysis.

\bibliographystyle{splncs04}
\bibliography{bibliography}

\end{document}